# Footprint-Driven Locomotion Composition

Christos Mousas[1], Paul Newbury[1], Christos-Nikolaos Anagnostopoulos[2]

[1]Department of Informatics, University of Sussex, Brighton BN1 9QJ, UK
[2]Department of Cultural Technology and Communication, University of the Aegean, University Hill, Mytilene 81100, Greece

## Abstract

*One of the most efficient ways of generating goal-directed walking motions is synthesising the final motion based on footprints. Nevertheless, current implementations have not examined the generation of continuous motion based on footprints, where different behaviours can be generated automatically. Therefore, in this paper a flexible approach for footprint-driven locomotion composition is presented. The presented solution is based on the ability to generate footprint-driven locomotion, with flexible features like jumping, running, and stair stepping. In addition, the presented system examines the ability of generating the desired motion of the character based on predefined footprint patterns that determine which behaviour should be performed. Finally, it is examined the generation of transition patterns based on the velocity of the root and the number of footsteps required to achieve the target behaviour smoothly and naturally.*

## 1. Introduction

During the past few years, the generation of motion sequences based on foot placement has been examined intensely [1][2][3][4][5]. This approach is important in cases where natural locomotion and collision avoidance is required, especially in highly constrained environments, such as indoors. In addition, since characters should be able to perform locomotion especially for goal-directed tasks, footprint-driven motion synthesis allows the generation of long motion sequences, in which the character can reach the exact desired position rather than an approximation. However, footprint-driven motion synthesis should not necessarily be limited to single actions, such as walking or running, as a vast number of applications related to virtual reality, such as video games, requires footprint-driven locomotion composition to involve multiple motion types.

The advantage of the presented solution, compared to existing footprint-driven motion synthesis techniques is the ability to generate a character's locomotion, rather than simple walking motions. This is achieved by extracting features related to footprint patterns to determine the different actions of the character during the composition, as well as by generating transition patterns based on the velocity of the character while another action is evolving. Therefore, in the presented solution, motion variations based on jumping, running, and stair stepping are presented to enhance the generated motion sequence. When dealing with different motions it is necessary to examine the generation of long motion sequences that retain the characteristics of each individual motion while allowing smooth transition between the different motion variations. Hence, the presented system automatically generates the desirable motion variations based on pre-computed properties of each motion and then disseminates these to the footprints resulting in smooth and continuous motion.

The rest of this paper is organised in the following sections. In Section 2 we discuss related work on motion synthesis techniques. In Section 3 the methodology of the proposed footprint-driven motion synthesis is presented. Section 4 deals with the locomotion composition process, where separate computation related to the motion patterns, behaviour transition process and improving continuity process are presented. In Section 5 the implementation of the proposed solution and the results obtained from the proposed motion synthesis process are presented. Finally, in Section 6 conclusions are drawn and potential future work is discussed.

## 2. RELATED WORK

Footprint-based methods for synthesising the locomotion of the characters are quite efficient. This is since, comparing both with the motion graphs and the search trees, footprint-based methods are responsible for driving the character in the exact desired position in the 3D space, rather than in an approximation. Specifically, the footprints are placed interactively by the user/developed in the 3D space. Each footprint denotes for animating virtual characters based on footprints that have been proposed the desired position and orientation of the character's feet while it contacts with the ground. Thus, while it is ensured the exact position of each characters foot in the 3D space, it is possible the character to follow a constrained path reaching the exact desired position. It should be noted that footprint-driven methods for animating the locomotion of a character are potentially the best way to describe the positioning of the motion data in the 3D space. Hence, during the past years various methodologies.

More specifically, Egges and Van Basten [1] proposed a data-driven method where a greedy nearest-neighbour approach warps the resulting animation to satisfy both spatial and temporal constraints. By using motion capture data as well, Wu et al. [2] proposed a procedural solution, in which the trajectory of the foot and the Center of Mass (CoM) are determined, thereby enriching the resulting motion, in accordance with an optimization method that enforces the resulting CoM to follow the desired trajectory. Another solution that uses motion capture data to generate the desired steps based on footprints was proposed by Huang and Kallmann [3], in which a pseudo-blending motion parameterisation process generates the desirable result. In this case, only the spatial position of the foot is achieved, rather than taking part in the motion synthesis process the exact foot positioning, as in the solution proposed by Van Basten et al. [4], where a hybrid interpolation scheme based on both orientation and Cartesian interpolation was used for synthesizing exact foot positioning. Finally, Choi et al. [5] proposed another solution that generates character's locomotion based on footprint. Specifically, in their method the virtual environment is sampled on possible footprint steps, constructing a roadmap. Then by augmenting the roadmap with a posture transition graph they traverse it to obtain the sequences of input motion clips.

On the other hand, physics and kinematics methods have been developed for animating virtual character based on footprints. Hence, Coros et al. [6] developed physical controllers that is aware of a footprint tries to follow it as closely as possible. Additionally, kinematics solutions proposed by Van de Panne [7] and Chung and Hahn [8]. Specifically, Van de Panne [7] proposed a procedural space-time work, in which the CoM trajectory is determined based on a physical optimizer according to the inverse kinematics determining leg motions. In the approach of Chung and Hahn [8] the motion synthesis is generated by a hierarchical system on top of footprints laid over uneven terrain.

The aforementioned methodologies, especially these related to data-driven, are able to provide quite reasonable results, especially in the part related to the naturalness of the synthesised motion. Generally, each of the aforementioned methods has its own advantages and disadvantages. For example, as mentioned in [2] the character is able to follow the footprints although, it is only ensured only the positional constraints of the foot. This limitation can easily be solved by enforcing kinematics constraints such as those used in [5]. Although, as kinematics are not always responsible for providing highly realistic posture of the character methodologies that are able to edit the motion data may be quite beneficial for solving the exact foot positioning problem. In [4] the exact foot positioning solved by proposing a methodology, which allows the motion data to be edited by fulfilling both the position and orientation constrains. In the methodology presented in this thesis a novel technique for solving this problem is introduced. Specifically, it is examined the ability of synthesising the required motion based on an optimization problem of the motion blending process. The result over the presented methodology ensures that the character fulfills both the position and orientation constraints, as well as that the computational cost that is required does not influences its real-time implementation.

Conversely, even if editing the motion data for fulfilling the necessary constraints, there are other issues that may influence the naturalness of the synthesised motion. For example, a character should not only being responsible for performing only single actions, such as regular walking motion. It should be able to perform different variations of motion sequences. Thus, methodologies that allow the synthesis of different behaviours should be examined. Based on [9] the incorporation of real-human measurements can be quite beneficial in ensuring the naturalness of the synthesised motion while different behaviours evolved in a locomotion sequence. Although, in [9] it was examined only the ability of a character to perform three different actions such as the walking, running and jumping. Specifically, in the aforementioned methodology, depending the target jump the character adjust its speed such as being able the system to provide a natural looking jump action of the character. A disadvantage of the aforementioned methodology is its inability to ensure a human like transition between the origin and the target speed. Thus, the resulted synthesised motion looks unrealistic. For that reason, the methodology presented in this thesis deals with the natural transitions that should evolve while a different target behaviour is required being synthesised by the system. For achieving this natural transition existing motion data were analysed. This analysis was performed on a basis of footsteps. Hence, by knowing in advance the number of steps where a character requires achieving a target behaviour it is possible to adjust progressively the speed of the character allowing a smooth and natural looking transition.

Finally, the interactivity between the user and such a system should be flexible enough, such as enabling almost every user to interact with the system allowing those to synthesise the required motion quite easily. In previous works, especially in [2] and [4], the user places the footprints in the 3D space and the system is responsible to synthesise the motion of a character. Although, the aforementioned methodologies are not able to synthesise different behaviours of the character. Moreover, it is not clear enough in the aforementioned methodologies what is the resulted motion of the character while a footprint in not located in a reachable area. Hence, in the presented methodology by introducing the footprint patterns the system is able to recognise and to synthesise different behaviours of the character. Moreover, for avoiding misconceptions of the system, by introducing some simple computations the system is able to ensure that each footprint lies in the action space of the existing motion data contained in the database, by placing additional footprints automatically.

## 3. METHODOLOGY

The motion synthesis process is split in to the three main components shown in the following subsections. These components are the ability to handle the motion capture data (Section 3.1), the motion extraction process (Section 3.2), and the motion blending process (Section 3.3) that is responsible for placing the foot in the exact required position.

### 3.1. Motion Handling

For each segmented motion the foot that performs the action is denoted as $p_{act}$, and the foot that supports the action as $p_{sup}$. Specifically, $p_{sup}$ is the foot that remains in contact with the ground during a single step, while the $p_{act}$ is the one that performs the step. In addition, for the foot performing the action a semantic label is assigned describing the starting and final position, such as $p_{start}$, which is the starting position, and $p_{end}$, which is the final position. Each supporting foot $p_{sup}$ is represented with the global position of foot joint, its $y - axis$ orientation, and the global position of the toe joint, such as $p_{sup} = (x^b, y^b, z^b, \theta^b, x^t, y^t, z^t)$ the positions of $p_{act}$ are represented in the same way but maintaining the local position, rather than the global, of the foot joint and toe joint, based on $p_{sup}$, that is, $p_{start} = (x_1^b, y_1^b, z_1^b, \theta_1^b, x_1^t, y_1^t, z_1^t)$ and $p_{end} = (x_2^b, y_2^b, z_2^b, \theta_2^b, x_2^t, y_2^t, z_2^t)$. Therefore, each motion contained in the database can be described as $m = \{p_{sup}, p_{start}, p_{end}, v_{root}, \xi\}$ where $v_{root}$ and $\xi$ are the velocity of the root and the semantic label of the motion respectively. Figure 1 depicts the parameters of a footstep.

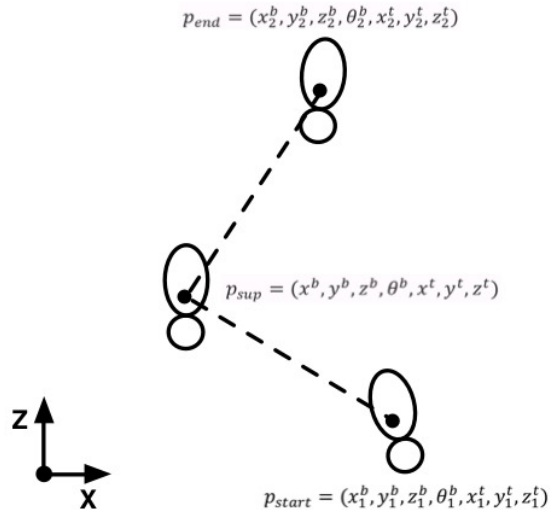

Figure 1. The parameters assigned for each step

For the extraction process, first all the supporting foot motions $p_{sup}$ must be aligned to the desired position and orientation of the footprint placed in the three dimensional space. This can be done by aligning the corresponding vectors. Then, the position of the supporting foot is represented as $f(p_{sup})$. Given $p_{start}$ and $p_{end}$ footprints, the next step is to extract the most suitable motions thereof for blending. Finally, it should be mentioned that the reason that the $p_{act}$ is assigned these parameters is the ability to extract and parameterise separately the reference motions of the foot joint and the toe joint. Then, those motions are assembled in order to provide the desirable result, as discussed in the following subsection.

### 3.2. Motion Extraction

For the motion extraction process, firstly each position of $p_{start}$ and $p_{end}$ with global orientation $\theta_i$ are sorted with the orientation of the desired reference position. Then, considering that $\theta_1 \leq \theta_i \leq \theta_n$, where $\theta_n$ is the $n-th$ registered orientation of the example motions, it is necessary to extract the most suitable of these, so as both the position and the orientation of the footprint lie within the desired position. For this reason, two example motion sequences are always considered, where $\theta_l \leq \theta_n$ must be satisfied for the first one, and $\theta_k \geq \theta_n$ for the second. Then, two more motions sequences are selected following the same procedure. Each footprint should be enclosed by two polygons (see Figure 2) that are produced based on the positions $p_{end}$ of the reference motion data, where each edge of the generated polygon corresponds to a $p_{end}$ reference position. Each suitable $p_{end}$ is used for the blending process. If multiple combinations of $p_{end}$ can satisfy this requirement, the polygon enclosing the foot joint, with the minimum sum of the distances is used for blending, thereby providing the desired result. The same process is used for extracting the most suitable motions that enclose the toe joint. It should be mentioned that the chosen motions sequences must fulfill both the positional and orientation criteria. This results in the ability to synthesise the exact foot placement.

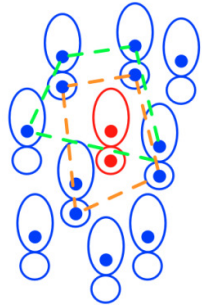

Figure 2. The polygons that enclose the foot joint (orange) and toe joint (green) are responsible for the blending process. The blue footprints denote the reference target motion sequences, and the red footprint a defined footprint.

### 3.3. Motion Blending

In this case, by directly blending the reference motions of the foot joint based on the form $P_{start}^f = \sum_{i=0}^{3} w_i P_i$, where $P_{start}^f$ is the desired position of the foot joint, $\sum_{i=0}^{3} w_i = 1$, and $P_i$ is the corresponding position of reference motion $m_i$, it is possible to retrieve the desired spatial position of the foot joint without ensuring the desired orientation. Then, using the same procedure, the most suitable motions that enclose the toe joint are blended in order to provide the desirable result.

For the motion blending process, having retrieved the blend weights $w_i$ ensuring the positional constraint of the foot joint and the toe joint, the next step is the recalculation of the weighted interpolation function of these for each degree of freedom separately, allowing the orientation constraints to be retrieved. In this case, the exact foot positioning can be computed as:

$$P(p_{foot}, p_{toe}) = v_1 \times P_{foot} + v_2 \times P_{toe} \qquad (1)$$

where $v_1$ and $v_2$ are the blendweight variable, $P_{foot}$ and $P_{toe}$ the position of the foot joint and the toe joint based on the resulted blending motion sequences. Although, since the direct blending does not ensures the exact foot positioning, the blendweights $v_1$ and $v_2$ are computed based in the following optimisation problem:

$$P(p_{foot}, p_{toe}) = min \sum_{i=0}^{S} \| v_1 \times P_{i,foot} + v_2 \times P_{i,toe} - P_i \| \qquad (2)$$

where $S$ denotes the total number of motion ($S = 8$, four for the foot and four for the toe) that are used for this blending process, $P_i$ is the $i-th$ example motion that used for the blending procrss, and $\sum_{i=1}^{2} v_i = 1$. Now, for solving this optimisation problem, the least square method was used. Thus, $v_1$ and $v_2$ are estimated as follows:

$$\begin{bmatrix} v_1 \\ v_2 \end{bmatrix} = \begin{bmatrix} P_{1,foot} & P_{1,toe} \\ ... & ... \\ P_{N,foot} & P_{N,toe} \\ 1 & 1 \end{bmatrix}^{+} \cdot \begin{bmatrix} P_{fp} \\ ... \\ P_{fp} \\ 1 \end{bmatrix} \qquad (3)$$

where $[]^{+}$ denotes the pseudo inverse of a matrix.

Using this approach, the character can satisfy more precisely the defined spatial and orientation constraints, as illustrated in Figure 3. Finally, it should be mentioned that during the blending process footsliding effect may appear. Hence, the technique proposed by Kovar et al. [10] that enforces an inverse kinematics solver for removing this undesired effect was enforced.

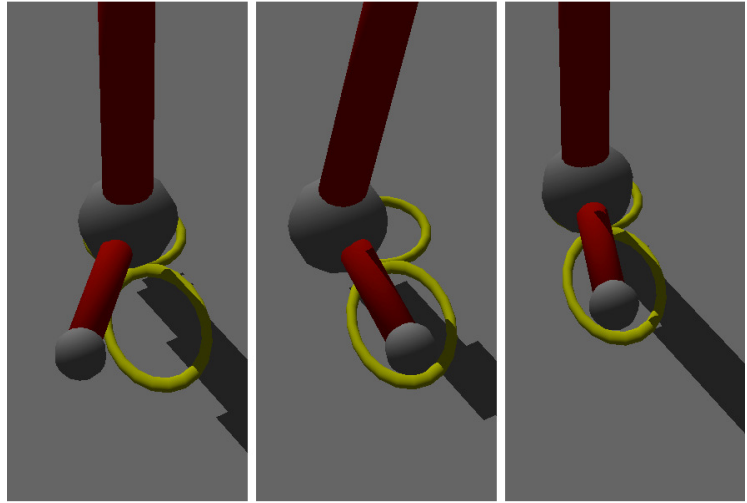

Figure 3. The spatial alignment of the foot based on the presented approach. The alignment of the foot joint (left), the alignment of the toe joint (middle) and the final generated alignment (right).

## 4. LOCOMOTION COMPOSITION

In the following subsections, the ability of synthesising the locomotion of the character is presented. More specifically, it is examined the ability of assigning the required behaviour of the character in the footprint patterns, and the ability of the transition process between different behaviours in the action transition graphs.

### 4.1. Footprint Patterns

In a simple footprint-driven walking sequence the character is able to perform the desired walking motion based on the actual motion sequences located in the database. However, while generating locomotion the actual step space can be extended. Hence, it is necessary for the system to recognise which locomotion should be generated based on the footprints placed in the three dimensional space. In general, changing the locomotion behaviour from walking to stair stepping is considered easy because it is possible to retrieve the foot placement based on its height. On the other hand, since the character should be able to exhibit more complicated behaviour, such as jumping, which could involve both feet jumping simultaneously or jumping by lifting one foot first, the system must be able to recognise which action matches the user inputs. Thus, even if the semantic labels of both actions are similar, the system should generate the most suitable motion.

In the presented approach, footprint patterns are generated to compute the relative position of the origin and target footprints. Thus, the footprint patterns are divided into two categories, namely, global footprint patterns, which are responsible for recognizing which action best matches the behaviours of walking, running and jumping, and local footprint patterns that are generated for jumping actions, so that the system can recognize which jumping action is the best match. As illustrated in Figure 4, for the global patterns among the three different behaviours, it was examined the distance between the two states of the action foot. Specifically, the plane $(x, z\ axies)$ distance between the two stances of $p_{act}$ is measured automatically by the system in a pre-processing stage by analysing the collection of motion clips for each particular behaviour. The maximum distance denotes the higher limit of the particular behaviour. Finally, it should be mentioned that when these limitations are based on a database of motion, as well as in cases where humans can perform the same motion with different variations, this approach cannot be generalised. Nevertheless, it can be used as a parameter to handle the motion capture data and generate the desired motions.

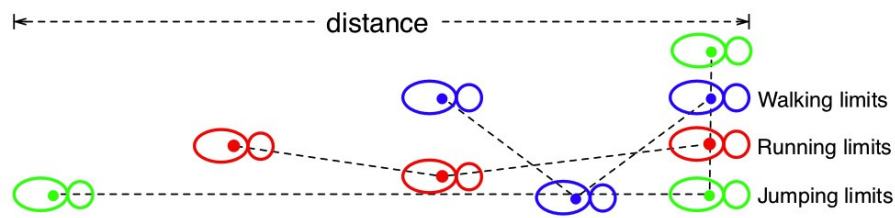

Figure 4. The global footprint patterns generated for walking, running and jumping behaviours.

The approach for generating local behaviours is quite important, especially when a jumping action needs to be performed, ensuring that the generated motion satisfies both the origin and the target footprints. Thus, four different local footprint patterns are generated (see Figure 5) allowing the system to automatically generate the desired jumping action that should be carried out. Specifically, for the generation process of the local footprint patterns, the system automatically analyses the angles $\theta_i$ and $\theta_j$ (as those angles are illustrated in Figure 6) between the feet positions both in lifting and in landing processes. This averaged angle is used as the threshold for determining the particular jumping action of the character. Thus, the character can perform jumping actions with either one or both feet lifting simultaneously, as well as with either one foot or both feet landing simultaneously. Finally, actions (b), (c), and (d) illustrated in Figure 5 are also mirrored for the other foot.

For the extraction process of walking, running and jumping, it is measured the distance that the active foot travels during a single step. Using this distance the system determines which behaviour to perform. If the extracted motion is either walking or running, the system directly builds the desired step. On the other hand, if the distance is within the limits of a jumping action, the system searches for the jumping behaviour that best matches the footprints. In this case, the generated angle $\theta$ between $p_{act}$ and $p_{sup}$ the original position, as well as at the target position, in conjunction to the distance between the $p_{sup}$ and $p_{act}$ are examined. A threshold angle, $\theta_{thres} \in [-\theta°, \theta°]$, which results from the jumping motion analysis process and a distance $d_i \leq d_{max}$ (see Figure 6) are used to generate the pattern in which both feet jump and/or land simultaneously.

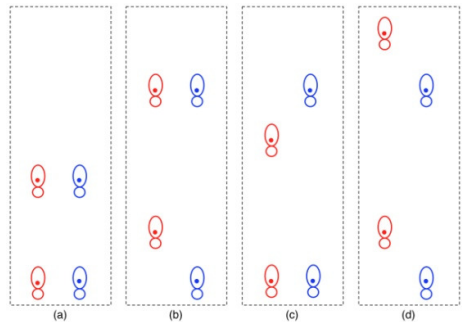

Figure 5. The four different footprint patterns used to generate jumping actions.

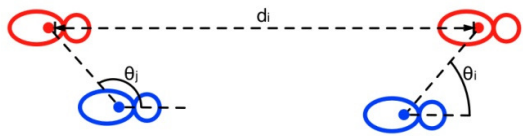

Figure 6. The parameters of the pattern in which both feet jump and/or land simultaneously.

To represent jumping actions, each pattern is assigned as $\Pi = \{\theta_i, \theta_j, d_i\}$, where $\theta_i$ and $\theta_j$ are the angles between $p_{sup}$ and $p_{act}$ at the lifting and landing position of the jumping action respectively, and depending on the actual value, the system measures one or both feet lifting or landing. $d_i$ denotes the actual distance between the $p_{start}$ and $p_{end}$ position of character's foot $p_{act}$ and it is used in order to extract jumping actions, in which the character travels a certain distance. Although, since during the jumping action there is not any foot contact with the ground the supporting foot does not exists. In this case, it should be mentioned that the system assigns as supporting foot the one which is used before the lifting process, as well as assigns as supporting foot the one which is first landed. Finally, it should be mentioned that the footprint patterns are common related with the existing motion sequences. Hence, in cases where a character with different height is calling to perform the desired locomotion, undesired actions may be occurred. An example where characters having a different high is illustrated in Figure 7.

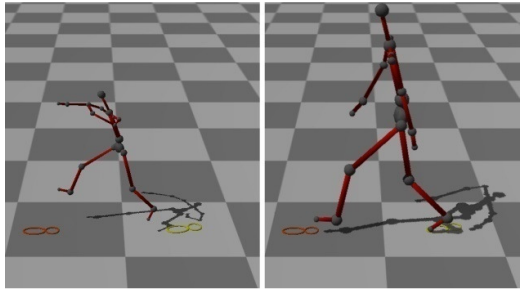

Figure 7. Characters with different high performs different motions while using the same footprints.

## 4.2. Behaviour Transition

This section examines the transition process improvement of the generated motion. This is because as the character changes behaviours, the synthesised motion should appear as natural as possible. As observed in the experiments, abruptmotions are generated especially when the character changes behaviour from, for example, running to stair stepping, walking to running and so on.

For achieving a smooth transition from one behaviour to another, a velocity-based approach for the root trajectory is used. To understand how each behaviour influences the generated motion sequence, in a pre-processing stage, the velocity of the character's root is measured while the character transits from one action to another. The results are semantically clustered based on the different behaviour transitions, linearly mapped according to the footsteps and the average number of footsteps required to proceed from one behaviour to any other is measured in conjunction with the velocity of the character's root. Transition graphs do not have to be generated for each combination of motions because motions that have a related velocity transit correctly, compared to those that the velocity changes dramatically, such as changing from walking to running actions. Hence, in the presented approach five different transition graphs are generated and used, as shown in Figure 8. Mirrored versions of these graphs are used to generate the inverse transition process. For retrieving those transition graphs the following steps, as those presented in the following subsections, were carried out. Finally, for a detailed explanation it is referred to read [12].

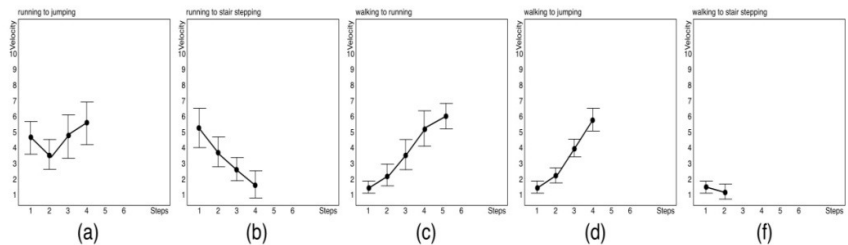

Figure 8. Transition graphs based on different transition behaviours: (a) from running to jumping (both feet jumping), (b) from running to stair stepping, (c) from walking to running, (d) from walking to jumping (one foot lifting and landing first), and (e) from walking to stair stepping. The horizontal axis shows the required footprints, while the vertical axis shows the velocity of the root in (m/s).

**4.2.1. Transition Identification**

For each motion, the velocity of the root, which characterises the transition process, was assigned as the main parameter for understanding the differences between the individual actions employed in the motion sequence. More specifically, the velocity of the root between two different actions should have a minimum or a maximum value that characterises one of the target actions. This approach was used to generate the desired velocity components assigned to each transition (see Table 1). For example, in the transition from a walking to running motion, the inverse transition, from running to walking, is executing by inversing the registered results, rather than using measurements from the motion capture data. The inverse process is used because it is assumed that this procedure can provide the desired result, as well as reduce the number of generated transition graphs.

Table 1. Discrete velocity characteristic used for the transition process between different actions. Velocity components with (-) were not computed.

| From \ To | Walking | Running | Jumping | Stair Stepping |
|-----------|---------|---------|---------|----------------|
| Walking   | -       | max     | max     | min            |
| Running   | -       | -       | max     | min            |

Based on Table 1, it is possible to generate the velocity that characterises each target action. However, as the velocity is not the sole parameter, the number of steps taken to move from the origin to the target is also measured. This measurement is made using a simple foot detector [11], which recognises foot contact with the ground based on the velocity and height of the foot. Using this approach, the velocity component that characterises the target action is determined, as is the time period for which the velocity of the action is maximised or minimised. It is then possible to determine the number of steps required to generate the velocity representing the maximum or minimum value.

**4.2.2. Transition Alignment**

Having calculated the root velocity in accordance with the foot contact approach, the next step of the method is to align the motions. The alignment process is based on the ability to align the number of steps with the root velocity. More specifically, two features are extracted for each motion: the number of steps $s_i$ and the velocity of the root $v_i$ at each step, such that each recorded motion $m_i$ is represented by $m_i = \{(s_1, v_1), \dots, (s_n, v_n)\}$, where $n$ denotes the $n - th$ registered step of the human derived from the motion capture data. Having extracted each action transition from those components, the next step is to align the motions (see Figure 9) based on the $i - th$ step, where the $i - th$ root velocity is equal to the maximum velocity $v_{max}$ or the minimum velocity $v_{min}$, depending on the discrete behaviour of the action transition mentioned above. This alignment ensures that all of the registered motions have the same transition behaviour and are aligned with the correct phase.

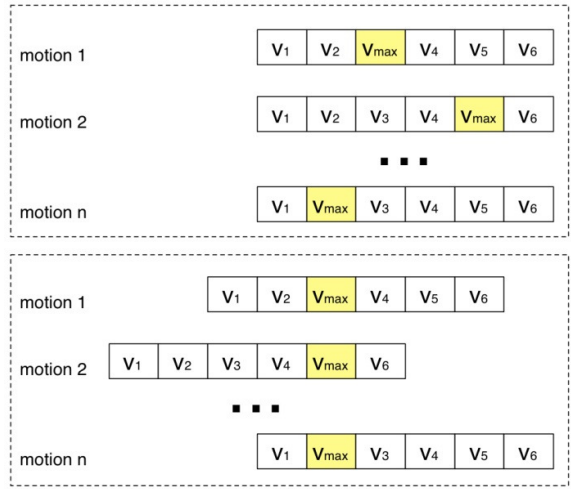

Figure 9. Alignment process of the step for which the velocity of the root is assigned to the velocity component that characterises the target action. An example is given for before the alignment (upper plate) and after the alignment (lower plate). Each rectangle denotes a step; the yellow rectangles denote the step at which the velocity component is executed.

### 4.2.3. Transition Graphs

The presented transition graphs represent the velocity of the root of each motion, as based on the corresponding steps. Considering that the number of steps for each registered motion varies, depending on the transition process, it is necessary to identify which steps should actually be included in the graphs. For the motion analysis process, it is necessary to remove the undesired steps (i.e., those that are not taking part of the transition process). Thus, for computing the actual steps that are required, the mean average of the number of steps before and after the velocity component is used to validate the number of required steps for the transition process.

Having aligned the steps based on the velocity property and having removed the undesired steps, to generate correct transition graphs, the actual numbers of steps and the root velocity are plotted. Based on the graph, it is possible to develop a general approach to the transition between two different actions by measuring the mean root velocity at each step. Based on the mean average of each action transition at each step, an action transition graph is constructed which presents the mean velocity of the root at each step of the character, presented as $m_{action} = \{(s_1, v_1^{mean}), ..., (s_n, v_n^{mean})\}$. With this representation of the transition process, it is possible to understand how each action transition evolves.

## 5. IMPROVING CONTINUITY

To improve the continuity of the movement, the system extracts information on a number of previous and subsequent footprints to determine which action should be performed. In the previous section generating motion that transits smoothly from one behaviour to another is discussed. The same metrics are used when a footprint is located in an area where the limits of two motions intersect in order to determine whether it denotes a long step or a running action. According to Figure 8(c) and considering that the original action is walking, the number of steps required by the character to generate a running action is four. Thus, by measuring only the previous footprints, the system cannot provide the exact desired motion since the subsequent footprints determine how the motion should evolve.

To enable the system to determine whether a running action should be performed, the next four steps are measured. Then, if all four steps are located within the intersection area of walking and running ($A_{walk} \cap A_{run}$), all of these steps are considered to be long steps. On the other hand, if at least one of those steps is located in the running action area only, the footprints are interpreted as a running action. In addition for each action the following limitation placed. The lower limit of the running actions is the upper limit of the walking actions. Similarly, the lower limit of the jumping actions is the upper limit of the running action. Hence, even if intersection between two actions exists, it is possible to extract the correct action of the character by using the distance limitation in conjunction with the information retrieved from the action transition graphs. It should be mentioned that the procedure between the walking and running action is used between running and jumping action. Even though this approach may not be the most efficient way of recognising which action best matches the footprints, it allows the system to determine and generate the different behaviours when the boundary areas intersect.

## 6. IMPLEMENTATION AND RESULTS

This section discusses the implementation of the presented solution, the system functionalities, and example motion sequences generated with the presented methodology. For the implementation of the presented solution, 100 steps of simple walking actions, 50 steps retrieved from running actions, 100 jumping actions (25 actions for each jumping condition) and 50 stair stepping actions were used. Mirrored versions were constructed for all the aforementioned motions, resulting in a total number of 600 motions. All motions were downsampled at 60 fps. Finally, the motion synthesis process were performed on an Intel i7 2.2 GHz processor with 8 GB RAM.

### 6.1. System Functionalities

Although the generated actions of a character are strictly limited to those contained in the database. A certain correction of those actions was necessary in order to avoid incorrect computations that are based on the ability of synthesising the locomotion of the character based on the motion handling approach. Depending on the action performed, two different approaches are implemented to correct the continuity. If a footprint is located higher in the three dimensional space than that supported by the registered motions in the database, the position of the supporting footprint $p_{sup}$ is changed by:

$$f(p_{sup}) = f(p_{sup}) + l_{stair}^{height} - f(p_{end}) \qquad (4)$$

where $l_{stair}^{height}$ is the upper limit of stair stepping motions. When the supporting footprint is also higher than the upper limit, the procedure that changes the height of the previous footprint continues iteratively backwards to the previous footprints. In this way, the approach provides an easy methodof generating multiple stair steps without aligning all the footprints.

On the other hand, automatic adjustments can be made in case a footprint is placed outside the limits of jumping actions, which have higher limits based on the global footprint patterns. In this case, the system divides the displacement between the end footprint $f(p_{end})$ and the upper limit for the previous footprints, that is:

$$f(p_i^{prev}) = f(p_i^{prev}) + \frac{\left| f(l_{action}^{height}) - f(p_{end}) \right|}{n_{prev}} \qquad (5)$$

where $f(p_i^{prev})$ is position of any previous footprint, $f(l_{action}^{height})$ is the position of the upper limit of the action used, and $n_{prev}$ is the total number of previous footprints. The function $f(\cdot)$ is used for describing the global position of the footprint. Although, this small adjustment can also be done in an arbitrarily large number of previous footprints. It was chosen to apply some simple rules, enabling the system to interpret easily any wrong position of a footprint.

By combining the two approaches, it is possible to enhance the synthesis process, avoid deadlock (e.g. a deadlock may occur while a footprint does not describe a particular action that the system could generate) and assistthe user in understanding to what extent each footprint influences the rest of the motion. Furthermore, the division process also works for stair stepping actions: if all previous footprints are deadlocked -that is all footprints are adjusted to higher limits- a new footprint is generated automatically on the top of the last one. With this approach it is ensured that always all the footprints that take part to the locomotion composition will be able to provide a valid motion. This is since all the footprints stay within the actual action space of motion where the character performs. Finally, it should be mentioned that imposing limitations directly on the footprints could circumvent these procedures, thereby avoiding footprints placed outside the boundary limits.

## 6.2. Results

For the completion of the computation process, each step required 0.189 seconds on average (based on both generated locomotion illustrated in Figure 13). The computational time required per step was divided between two different processes. Motion parameterisation, and motion blending, took 39%. The motion synthesis process based on motion graphs [96] used the remaining computation time, that is, 61%. The computational cost of the method may seem high, but given the complexity of the problem and compared to other solutions presented in the literature, it undoubtedly represents an improvement to existing systems. Figure 10 illustrates the linkage between the number of motion sequences that were used, in conjunction with the ability of synthesizing each single step of the character. Considering the computational cost of the solution proposed by Egges and Van Basten [1] (where they use an Intel Pentium Centrino 2.4 GHz, and 200 motion sequences stored in their database), a single step requires 3ms computational time on average. The computational time of the presented approach, for 200 motion sequence (100 walking, 50 running, 30 stair stepping, and 20 jumping) was estimated to be 0.076 seconds. The cost difference between the solution proposed by Egges and Van Basten[1] was anticipated, since in the presented approach additional parameters related to the motion extraction process, such as action transition and exact foot positioning, influence the motion synthesis process.

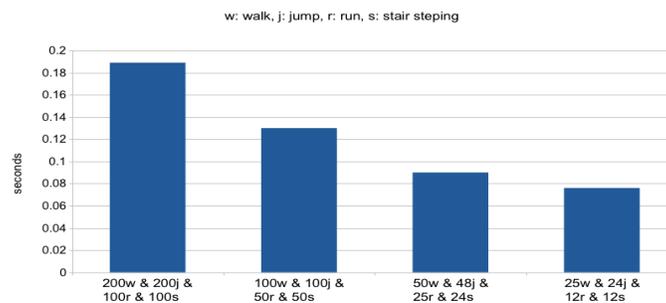

Figure 10. The time required for the generation of a single step, based on different database size.

The innovation of the presented solution is the ability to infer and synthesize the correct behaviour of the character based on the defined footprints. Specifically, the system is able to generate jumping actions based on the patterns presented earlier. Figure 11 shows the generation of different jumping actions based on the placement of the footprints. In addition, the ability of the locomotion composition process to generate long motion sequences with the footprints located across a desired path, which can be either straight or curved, was tested. Figure 12 illustrates example motions generated based on the different paths. It was also examined the ability of the presented approach to generate different behaviours during the locomotion. Figure 13 depicts two long motion sequences in which all the examined behaviours are implemented. Finally, it should be mentioned that even if a large collection of motion sequences is used to generate the motion of the virtual character, there are still motions that the database does not contain. In this case, the movement generated does not appear to be completely natural, since the motion blending technique cannot always provide the desired results.

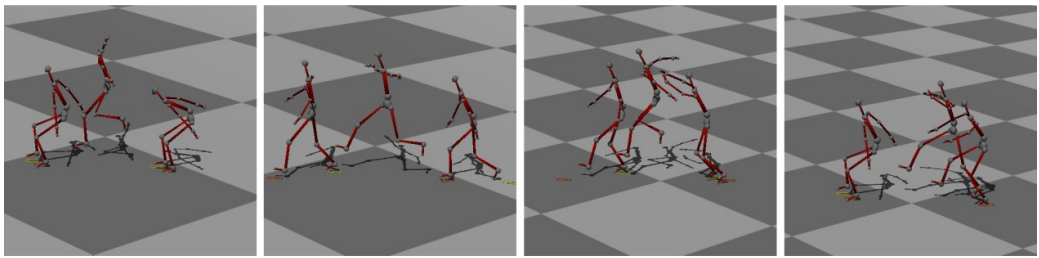

Figure 11. Generated jumping motions based on different footprint patterns.

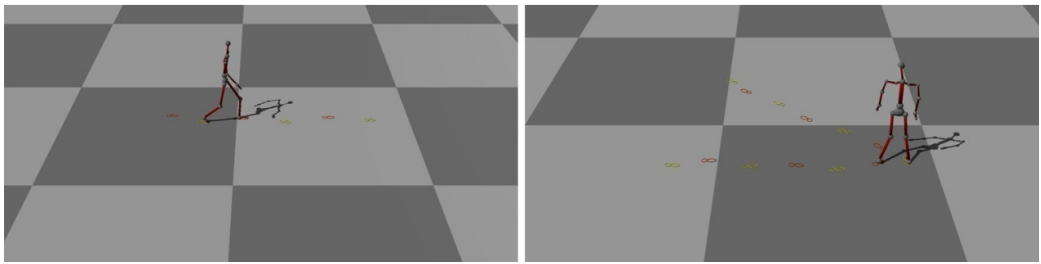

Figure 12. Generated motion based on a straight (left) and curvature (right) path.

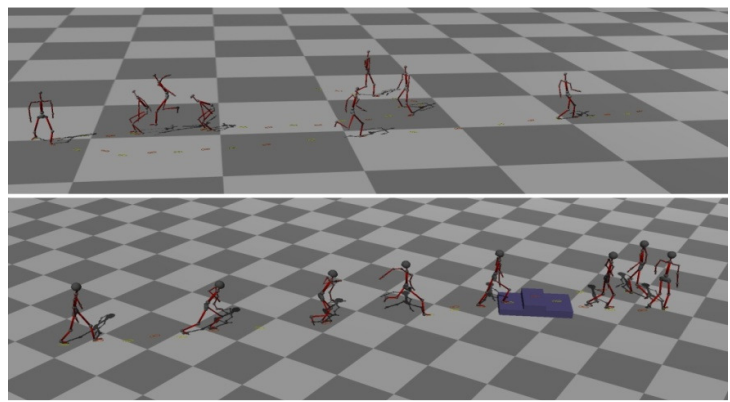

Figure 13. Different behaviours implemented in a single locomotion procedure. Example motions that consist of 46 footprints with walking, sidestep, running and jumping actions (upper row), and another

generated locomotion procedure that consists of 24 steps with walking, running, jumping and stairs
stepping actions (lower row).

## 7. DISCUSSION

In this paper a solution for automaticallysynthesizing continuous locomotion of a virtual character
based on footprints was presented. In particular, it is showed that the presented solution could
generate a vast number of different actions, such as running and jumping actions. Since the
system is able to recognise the desired actions of the character, the presented solution adds the
ability of synthesising the desired action of the character by assigning the motion synthesis
process to footprint patterns, enabling the identification of each action. In addition, especially in
common related actions, such as the jumping actions, local footprint patterns developed to
determine which motion is the best match. Additionally, it is often desirable to generate not only
continuous motion based on different behaviours, but also continuous motion in which the
different behaviours can transit smoothly, it has been shown that generating motion transition
patterns related to the velocity of the root in accordance with the required number of footprints
result in a smooth transition from one behaviour to another. Although, it should be mentioned that
the effectiveness of the method is highly depended on the number of motion sequences. Thus, the
more variety of reference motion sequences, the more variety of behaviours of the character that
can be generated.Finally, in a future implementation we would try to examine the ability of
integrating the footprint-based motion synthesis methodology to a recently presented path
planning methodology [13][14] for simulating the human locomotion in highly constrained virtual
environments. Thus, we assumethat a high quality humanlike space-time motion planningmodel
could be generated.

## Authors

**Christos Mousas** holds a BSc in Audio-Visual Science and Art from Ionian University (Greece), and an MSc in Multimedia Applications and Virtual Environments from the University of Sussex (UK). Currently, he is a PhDcandidate at the Department of Informatics at the University of Sussex. His main research interests include interactive techniques for editing and synthesis of motion capture data for virtual characters, virtual reality, and virtual humans.

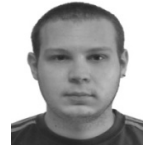

**Christos-NikolaosAnagnostopoulos** was born in Athens, Greece in 1975. He received his Mechanical Engineering Diploma from the National Technical University of Athens (NTUA) in 1998, and the Ph.D. degree from the Electrical and Computer Engineering Dpt., NTUA in 2002. From 2008, he serves the University of the Aegean as Assistant Professor in the Cultural Technology and Communication Department. Dr.Christos–NikolaosAnagnostopoulos is a member of the Greek chamber of Engineers and member of IEEE. His research interests are image processing, computer vision, neural networks and artificial intelligence. He has published more than 100 papers in journals and conferences, in the above subjects as well as other related fields in informatics.

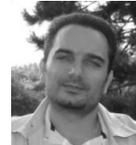

**Paul Newbury**received his BSc (Hons) in Electronic Engineering in 1991, MSc in Intelligent Systems in 1993 and a PhD in Computer Science in 1999. He is currently a member of the Centre for Computer Graphics at the University of Sussex, a Senior Lecturer in Multimedia Systems and has published articles in image processing, data compression, computer graphics, virtual and augmented reality and simulation of pervasive environments.

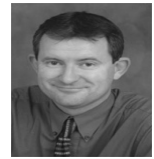